\documentstyle[epsf,aps]{revtex}
\draft
\begin{document}
\title{Energies of the ground state and first excited $0^{+}$ state\\ 
in an exactly solvable pairing model}
\author{N. Dinh Dang}
\address{RI-beam factory project office, RIKEN, 2-1 Hirosawa, Wako, 
351-0198 Saitama, Japan\\
and\\
Institute for Nuclear Science and Technique, VAEC, Hanoi, Vietnam
}
\date{\today}
\maketitle
\begin{abstract}
Several approximations are tested by calculating 
the ground-state energy and the energy of the 
first excited $0^{+}$ state  using an exactly solvable model with 
two symmetric levels interacting via a pairing force. 
They are the 
BCS approximation (BCS), Lipkin - Nogami (LN) method, random-phase 
approximation (RPA), quasiparticle RPA (QRPA), the renormalized RPA (RRPA), 
and renormalized QRPA (RQRPA). It is shown that, in the strong-coupling 
regime, the QRPA which neglects the scattering term of the model 
Hamiltonian offers the best fit to the exact solutions.
A recipe is proposed using the RRPA and RQRPA in combination with the 
pairing gap given by the LN method. Applying this recipe, it is shown that 
the normal-superfluid phase transition is avoided, and a reasonably
good description for both of the
ground-state energy and the energy of the first excited $0^{+}$ state is 
achieved.
\end{abstract}
\section{Introduction}
The random-phase approximation (RPA)~\cite{Rowe,Ring} 
is one of the most popular method 
in the theoretical microscopic study of nuclear structure. It includes
the correlations beyond the mean field models such as the Hartree-Fock 
(HF) approximation with a phenomenological interaction, e.g. the Skyrme 
interaction. Being also a computationally simple method, the RPA 
serves as a powerful tool in treating all the 
excited states in nuclei, which are beyond the 
reach of the full diagonalization within a shell-model basis.
The quasiparticle RPA (QRPA), in which the quasiparticle correlations and
excitations are considered, has shown its prominent role in treating the
open-shell nuclei, where the superfluid pairing correlations are important.

Recent developments in nuclear structure including 
$\beta$-decay and double-beta ($\beta\beta$) -decay physics, the study of various 
types of giant resonances, and 
the prospect of using radioactive beams to study nuclei far from the line of 
$\beta$-stability have sparked off a renewed interest 
in the efficiency of the RPA compared to other microscopic calculations
~\cite{testRPA}. More refined and exact treatments of the pairing problem in 
nuclei have also been proposed in~\cite{pairing}.

Recently the accuracy of the RPA in describing the binding energies 
has been tested 
using the HF+RPA calculations within schematic exactly solvable 
models~\cite{Hagino1,Hagino} as 
well as for nuclei throughout the $sd$-shell and the lower $pf$-shell
~\cite{Stetcu}.
The tests using exactly solvable Lipkin models~\cite{Hagino1,Hagino} 
has shown that the HF+RPA (QRPA) 
calculations yield a very good ground state energy
except for the region of the pairing interaction around the point
where the pairing gap collapses. Mean while, the tests using more 
realistic shell-model Hamiltonian have pointed out that
the binding energy predicted by RPA is generally, but not always, 
satisfactory~\cite{Stetcu,Ullah}. A number of 
suggestions have been made to improve the reliability 
of the RPA. They can be classified into two groups. 
The first group
includes the approaches to improve the treatment of pairing correlations. 
Among them are the Hartree-Fock-Bogolyubov (HFB) + QRPA, the number 
projection~\cite{Ring}, the variational method using boson 
expansion~\cite{Sambataro}, 
and the Lipkin-Nogami (LN) method~\cite{LN1,LN}. The LN method is an 
approximation to remove the fluctuations due to the violation
of particle number conservation of the BCS approximation, which leads to the 
collapses of the pairing gap at a certain critical value of the pairing 
interaction.
The second group includes various treatments of the ground-state 
correlations beyond RPA, which occur due to the violation of the Pauli 
principle when the quasiboson approximation
is used within the RPA or QRPA.   
In a way similar to the particle number violation within the BCS approximation, 
the violation of the Pauli principle by treating the fermion pairs as 
bosons within the quasiboson approximation 
leads to the collapse of the RPA at a critical value 
of the interaction parameter. Various approaches have been proposed to 
renormalize the RPA (QRPA) to remove this inconsistency
~\cite{Hara,Klein,Schuck,Krmpotic,RRPA,MRPA}. 

Since the RPA, first of all, is a theory of excited states, 
an improvement of the accuracy of 
the RPA should give, as a first step, 
a better description for both of the energies of the 
ground state and the first excited state simultaneously.
In the present paper, we are going to propose a recipe for such 
improvement using existing approximation schemes, namely the LN method
for the pairing gap and the renormalized RPA (renormalized QRPA) to treat the ground 
state correlations beyond RPA. In order to compare the result with the 
exact solution we limit ourselves to the study of the ground state and the 
first excited $0^{+}$ state within an well-known
exactly solvable two-level model, which was introduced for the first time in 
Ref. \cite{Hogaasen}, and widely used in the study 
of pairing correlations.    
 
The paper is organized as follows. The pairing Hamiltonian applied to 
the two-level model is discussed in 
Sec. I. The BCS approximation, the LN method, and the results obtained 
within these approximations for the pairing gap in 
the two-level model under consideration 
are summarized in Sec. II.
The RPA, QRPA and their renormalized versions are presented in Sec. III, 
where the results for the ground-state energy and energy of the first 
excited $0^{+}$ state are analysed. Also discussed in the same section are 
results obtained following a recipe 
which combines the LN-method and the renormalized QRPA 
(renormalized RPA).
The last section summarizes the paper, where conclusions are drawn.
\section{Pairing Hamiltonian}
We consider the well-known pairing Hamiltonian
of nucleons interacting via a pairing force between the time-conjugate 
orbitals with angular-momentum quantum number $j$ 
\begin{equation}
H=\sum_{j}\epsilon_{j}N_{j}-G\sum_{jj'}\sqrt{\Omega_{j}}\sqrt{\Omega_{j'}}
A^{\dagger}_{j}A_{j'}~,
\label{H}
\end{equation}
where $\epsilon_{j}$ are the single-particle energies, and $G$ is the pairing 
constant. The nucleon number
operator for the $j$ shell  is
\begin{equation}
N_{j}=\sum_{m}a^{\dagger}_{jm}a_{jm}~.
\label{N}
\end{equation}
The pairing operators $A^{\dagger}_{j}$ and $A_{j}$ are given by
\begin{equation}
A_{j}^{\dagger}=\frac{1}{\sqrt{\Omega_{j}}}\sum_{m>0}
a_{jm}^{\dagger}a^{\dagger}_{j\widetilde{m}}~,\hspace{5mm} 
A_{j}=(A^{\dagger}_{j})^{\dagger}~,
\label{A}
\end{equation}
where the tilde denotes the time-reversal operation, e.g. 
$a^{\dagger}_{j\widetilde{m}}=(-1)^{j-m}a^{\dagger}_{j-m}$.
These operators satisfy the following commutation relations:
\begin{equation}
[A_{j},A^{\dagger}_{j'}]=\delta_{jj'}\bigg(1-\frac{N_{j}}{\Omega{j}}\bigg)~,
\label{AA}
\end{equation}
\begin{equation}
[N_{j},A^{\dagger}_{j'}]=2\delta_{jj'}A^{\dagger}_{j}~,\hspace{6mm} 
[N_{j},A_{j'}]=-2\delta_{jj'}A_{j}~.
\label{NA}
\end{equation}
The exact solutions of this pairing Hamiltonian have been found
in Ref. \cite{Richarson} and known as the Richarson's solution.
In the present paper, we consider only 
a simple schematic model. 
It has $N$ particles occupying two levels with the same shell degeneracy 
$\Omega=j+1/2$ so that $\Omega=N/2$.
The upper level simulates the degenerates states with energy $\epsilon/2$ 
and magnetic quantum 
numbers $m$ (0$<m\leq\Omega$), while the lower level is for the states 
with energy $-\epsilon/2$ and the magnetic quantum numbers $-m$. 
The distance between two levels is, therefore, $\epsilon$, which 
will take the value equal to 1 MeV in all calculations in this paper.
The exact solution in this case is easy to be found  using the 
SU(2) algebra, which the operators $J_{+}=\sqrt{\Omega_{j}}A^{\dagger}_{j}$,
$J_{-}=(J_{+})^{\dagger}$, and $J_{0}=(N_{j}-\Omega_{j})/2$ obey. This leads
to the diagonalization of the Hamiltonian (\ref{H}), whose matrix elements 
are
\[
\langle J,M_{1}',M_{2}'|H|J,M_{1},M_{2}\rangle=
-\epsilon(M_{1}-M_{2})\delta_{M_{1},M_{1}'}\delta_{M_{2},M_{2}'}
\]
\[
-G[2J(J+1)-M_{1}(M_{1}-1)-M_{2}(M_{2}-1)]
\delta_{M_{1},M_{1}'}\delta_{M_{2},M_{2}'}
\]
\begin{equation}
-G\sqrt{J(J+1)-M_{1}(M_{1}-1)}\sqrt{J(J+1)-M_{2}(M_{2}+1)}
\delta_{M_{1},M_{1}'+1}\delta_{M_{2},M_{2}'-1}
\label{me}
\end{equation}
\[
-G\sqrt{J(J+1)-M_{1}(M_{1}+1)}\sqrt{J(J+1)-M_{2}(M_{2}-1)}
\delta_{M_{1},M_{1}'-1}\delta_{M_{2},M_{2}'+1}~,
\]
with $J=\Omega/2$ and $-J\leq M_{i}\leq J$ ($i=$ 1, 2). 
Among the obtained eigenenergies ${\cal E}_{i}$, the lowest one, 
${\cal E}_{0}$, is the exact ground-state energy, while the exact energy 
$\omega_{\rm ex}$ of 
the first excited  $0^{+}$ state is 
$
\omega_{\rm ex}={\cal E}_{1}-{\cal E}_{0}~.
$

It is convenient to study a fermion system with superfluid pairing
using the Bogolyubov transformation from particle operators
$a_{jm}^{\dagger}$ and $a_{jm}$ to the quasiparticle ones, 
$\alpha_{jm}^{\dagger}$ and $\alpha_{jm}$. 
The quasiparticle representation of the Hamiltonian
(\ref{H})  is given in \cite{Hogaasen,Dang}, which we quote here again
for the convenience in further discussions:
\[
H=a+\sum_{j}b_{j}{\cal N}_{j}+\sum_{j}
c_{j}({\cal A}_{j}^{\dagger}+{\cal A}_{j})
+\sum_{jj'}d_{jj'}{\cal A}_{j}^{\dagger}
{\cal A}_{j'}
\]
\begin{equation}
+\sum_{jj'}g_{j}(j')({\cal A}_{j'}^{\dagger}{\cal N}_{j}
+{\cal N}_{j}{\cal A}_{j'})
+\sum_{jj'}h_{jj'}
({\cal A}_{j}^{\dagger}{\cal A}_{j'}^{\dagger}
+{\cal A}_{j'}{\cal A}_{j})+\sum_{jj'}q_{jj'}{\cal N}_{j}{\cal N}_{j'}~.
\label{qH}
\end{equation}
Here ${\cal N}_{j}$ is the operator of the 
quasiparticle number on the $j$-shell, while
${\cal A}_{j}^{\dagger}$ and ${\cal A}_{j}$ are the creation and destruction
operators of a pair of time-conjugate quasiparticles:
\begin{equation}
{\cal N}_{j}=\sum_{m}\alpha^{\dagger}_{jm}\alpha_{jm}~,\hspace{5mm} 
{\cal A}_{j}^{\dagger}=\frac{1}{\sqrt{\Omega_{j}}}\sum_{m>0}
\alpha_{jm}^{\dagger}\alpha^{\dagger}_{j\widetilde{m}}~,\hspace{5mm} 
{\cal A}_{j}=({\cal A}^{\dagger}_{j})^{\dagger}~.
\label{calA}
\end{equation}
Their commutation relations are similar to those for nucleon operators in
Eqs. (\ref{AA}) and (\ref{NA}), namely:
\begin{equation}
[{\cal A}_{j},{\cal A}^{\dagger}_{j'}]=\delta_{jj'}(1-\frac{{\cal N}_{j}}{\Omega{j}})~,
\label{calAA}
\end{equation}
\begin{equation}
[{\cal N}_{j},{\cal A}^{\dagger}_{j'}]=2\delta_{jj'}{\cal A}^{\dagger}_{j}~,\hspace{6mm} 
[{\cal N}_{j},{\cal A}_{j'}]=-2\delta_{jj'}{\cal A}_{j}~.
\label{calNA}
\end{equation}
The coefficients in Eq. (\ref{qH}) are
\begin{equation}
a=2\sum_{j}\Omega_{j}\epsilon_{j}v_{j}^{2}
-G(\sum_{j}\Omega_{j}u_{j}v_{j})^{2}-G\sum_{j}\Omega_{j}v_{j}^{4}~,
\label{a}
\end{equation}
\begin{equation}
b_{j}=\epsilon_{j}(u_{j}^{2}-v_{j}^{2})+2Gu_{j}v_{j}\sum_{j'}\Omega_{j'}u_{j'}v_{j'}
+Gv_{j}^{4}~,
\label{b}
\end{equation}
\begin{equation}
c_{j}=2\sqrt{\Omega_{j}}\epsilon_{j}u_{j}v_{j}-G\sqrt{\Omega_{j}}
(u_{j}^{2}-v_{j}^{2})\sum_{j'}\sqrt{\Omega_{j'}}u_{j'}v_{j'}-
2G\sqrt{\Omega_{j}}u_{j}v_{j}^{3}~,
\label{c}
\end{equation}
\begin{equation}
d_{jj'}=-G\sqrt{\Omega_{j}\Omega_{j'}}(u_{j}^{2}u_{j'}^{2}+v_{j}^{2}v_{j'}^{2})=d_{j'j}~,
\label{d}
\end{equation}
\begin{equation}
g_{j}(j')=Gu_{j}v_{j}\sqrt{\Omega_{j'}}(u_{j'}^{2}-v_{j'}^{2})~,
\label{g}
\end{equation}
\begin{equation}
h_{jj'}=\frac{G}{2}\sqrt{\Omega_{j}\Omega_{j'}}
(u_{j}^{2}v_{j'}^{2}+v_{j}^{2}u_{j'}^{2})=h_{j'j}~,
\label{h}
\end{equation}
\begin{equation}
q_{jj'}=-Gu_{j}v_{j}u_{j'}v_{j'}=q_{j'j}~,
\label{q}
\end{equation}
where $u_{j}$ and $v_{j}$ are the coefficients of the Bogolyubov 
transformation. Hereafter we will refer to the terms at the right-hand side
(RHS) of Eq. (\ref{qH}), which contain the coefficients $a_{j}$ , $b_{j}$, 
etc. as the a-term, b-term, etc., respectively.

\section{Gap equations}
\subsection{BCS approximation}
The BCS equation is usually obtained using the variational
procedure to get the minimum of the average value of $H-\lambda\hat{N}$ 
($\lambda$ is the chemical potential, $\hat{N}=\sum_{j}N_{j}$ is the 
particle-number operator) over
the BCS ground state taken as the quasiparticle vacuum $|0\rangle_{\alpha}$, 
i.e. $\alpha_{jm}|0\rangle_{\alpha}=$ 0, where $
|0\rangle_{\alpha}=\prod_{j,m>0}(u_{j}+v_{j}a_{jm}^{\dagger}a_{j\widetilde{m}}
^{\dagger})|0\rangle$ with $a_{jm}|0\rangle=$ 0.
Within the BCS approximation, only the a-term in Eq. 
(\ref{qH}) contributes, which leads to the well-known BCS equation to 
determine the gap $\Delta$ and chemical potential $\lambda$:
\begin{equation}
\Delta=G\sum_{j}\Omega_{j}u_{j}v_{j}~,\hspace{5mm}
N=\sum_{j}\Omega_{j}\bigg(1-\frac{{\epsilon_{j}'}-\lambda}{E_{j}}\bigg)~,
\label{BCS}
\end{equation}
where the single-particle energy is 
$\epsilon_{j}'=\epsilon_{j}$ if 
the self-energy term $-Gv_{j}^{2}$ is neglected, or 
$\epsilon_{j}'=\epsilon_{j}-Gv_{j}^{2}$ 
if the self-energy term is included. 
The quasiparticle energy is 
$E_{j}=\sqrt{(\epsilon_{j}'-\lambda)^{2}+\Delta^{2}}$. The
$u_{j}$ and $v_{j}$ 
coefficients are given as
\begin{equation}
u_{j}^{2}=\frac{1}{2}\bigg(1+\frac{\epsilon'_{j}-\lambda}{E_{j}}\bigg)~,
\hspace{5mm}
v_{j}^{2}=\frac{1}{2}\bigg(1-\frac{\epsilon'_{j}-\lambda}{E_{j}}\bigg)~.
\label{ujvj}
\end{equation}
The a-term (Eq. (\ref{a})) is actually the ground state energy within the BCS 
approximation, since
this is the only term that remains in the average over the quasiparticle 
vacuum $|0\rangle_{\alpha}$, where the second term 
can be now replaced with $-\Delta^{2}/G$ using the 
BCS equation (\ref{BCS}). 

In the present two-level model, using the BCS equations (\ref{BCS}) and 
the property $u_{i}^{2}+v_{i}^{2}=$ 1 with $i=$ 1 (lower level) or 2 (upper 
level), it is easy to see that the quasiparticle energy $E$ and the 
chemical potential $\lambda$ are state-independent, namely  
\begin{equation}
E_{1}=E_{2}=E=G\Omega~,\hspace{8mm} 
\lambda=-\frac{G}{2}~.
\label{lambda}
\end{equation}
The gap $\Delta$ and the $u$ and $v$ coefficients are~\cite{Hagino}
\begin{equation}
\Delta=G\Omega\sqrt{1-\bigg[\frac{G_{\rm cr}}{G}\bigg]^{2}}~,
\label{gap}
\end{equation}
\begin{equation}
u_{1}^{2}=v_{2}^{2}=\frac{1}{2}\bigg(1-\frac{\widetilde{\epsilon}}{2G\Omega}
\bigg)=u^{2},~
\hspace{5mm} 
v_{1}^{2}=u_{2}^{2}=\frac{1}{2}\bigg(1+
\frac{\widetilde{\epsilon}}{2G\Omega}\bigg)=v^{2}~,
\label{uv}
\end{equation}
where
\begin{equation}
\widetilde{\epsilon}=
\left\{\begin{array}{ll}
\epsilon & \textrm{neglecting the self-energy term~,}\\\\
{2\Omega}\epsilon/({2\Omega-1}) & \textrm{including the self-energy 
term~,}
\end{array}\right.
\label{epst}
\end{equation}
and
\begin{equation}
G_{\rm cr}=\frac{\widetilde{\epsilon}}{2\Omega}~.
\label{Gcrit}
\end{equation}
The BCS ground state energy in this model becomes
\[E_{\rm BCS}=\Omega\epsilon(1-2v^{2})-\Delta^{2}/G-G\Omega(u^{4}+v^{4})
\]
\begin{equation}
=\left\{\begin{array}{ll}
-\epsilon^{2}/2G-\Delta^{2}(1-1/2\Omega)/G-G\Omega& \textrm{neglecting 
the self-energy term~,}\\\\
-\epsilon^{2}/(2G(1-1/2\Omega))-\Delta^{2}(1-1/2\Omega)/G-G\Omega~~~
& \textrm{self-energy term included~.}\end{array}\right.
\label{EBCS}
\end{equation}
This shows that the effect of the self-energy term on the ground-state energy 
becomes negligible at a large particle number when $1/2\Omega\ll$ 1.
The major drawback of the BCS method is that the 
BCS wave function is not an eigenstate of the particle 
number operator $\hat{N}$. Therefore the particle-number
fluctuations 
$\Delta N^{2}={~}_{\alpha}\langle 0|\hat{N}^{2}|0\rangle_{\alpha} 
-{~}_{\alpha}\langle| 0\hat{N}|0\rangle_{\alpha}^{2}
=4\sum_{j}\Omega_{j}(u_{j}v_{j})^{2}=2\Delta^{2}/{G^{2}\Omega}$~\cite{Dang}
cause the 
inaccuracy of this method. 
Using the exact commutation relation (\ref{calAA}), 
and ${}_{\alpha}\langle 0|{\cal N}_{j}|0\rangle_{\alpha}=$ 0, we see
that 
\begin{equation}
{~}_{\alpha}\langle 0|[{\cal A}_{j},{\cal 
A}^{\dagger}_{j'}]|0\rangle_{\alpha}=\delta_{jj'}~,\hspace{5mm} 
{~}_{\alpha}\langle 0|[{\cal N}_{j},{\cal 
A}^{\dagger}_{j'}]|0\rangle_{\alpha}=0~,
\label{calA0}
\end{equation}
which imply that, within the BCS approximation, the quasiparticle-pair 
operators ${\cal A}^{\dagger}_{j}$ and ${\cal A}_{j}$ behave like 
bosons (the so-called Cooper pairs). Such
violation of the Pauli principle between quasiparticles has 
the same origin as the quasiboson approximation 
used in the RPA, which will be discussed in the next 
section. This 
leads to the collapse of the pairing gap at the critical value $G_{\rm cr}$, 
below which the BCS equations (\ref{BCS}) yields the imaginary solution.
In the present two-level model, $G_{\rm cr}$ is given by Eq. (\ref{Gcrit}).
Such kind of 
critical behavior inspired a speculation of the existence of a phase 
transition from the normal phase, where the gap is absent, to the 
superfluid phase with a nonzero gap. In the absence of the pairing gap, 
only the sum over the hole ($h$) 
state remains in expression for the 
ground-state energy (\ref{a}), as $v_{h}=$ 1 and $v_{p}=$ 0. This gives 
the HF ground-state energy within the present two-level model as
\begin{equation}
E_{\rm HF}=\langle{\rm HF}|H|{\rm HF}\rangle=2\sum_{j_{h}}\Omega_{j_{h}}\epsilon_{j_{h}}-G\sum_{j_{h}}
\Omega_{j_{h}}=-\Omega(\epsilon+G)~,
\label{EHF}
\end{equation}
where $|{\rm HF}\rangle=\prod_{j}a^{\dagger}_{jm}|0\rangle$ is the HF ground 
state.  
However, the superfluid-normal phase transition in a system with a finite 
particle number is spurious as it does not exist in 
the exact calculations~\cite{Rho} as well as in the method using particle-number 
projection~\cite{Ring}, where the gap is finite at all finite values of 
$G$~\cite{Ring}.
Since carrying out the particle-number projection in calculations using 
realistic spectra is numerically complicate, a simple approximate number 
projection has been proposed, 
which is known as the Lipkin-Nogami (LN) method~\cite{LN1,LN} and summarized 
below for the present two-level model. 
\subsection{Lipkin-Nogami (LN) method}
The LN method 
has gained a great popularity as it provides a simple and computationally
easy way to go beyond the pairing mean-field.
Within this method, the particle-number fluctuations are removed by
adding the term $\lambda_{2}\hat{N}^{2}$, and carrying out the variational 
procedure over the average value of $H-\lambda\hat{N}-\lambda_{2}\hat{N}^{2}$
in the quasiparticle vacuum $|0\rangle_{\alpha}$. Details of this method is 
given in Ref. \cite{LN}. For the present two-level model, it gives
\begin{equation}
\Delta_{\rm LN}=G\Omega\sqrt{1-\frac{\bar{\epsilon}^{2}}{4G^{2}\Omega^{2}}}~,
\label{LNgap}
\end{equation}
\begin{equation}
u^{2}=\frac{1}{2}\bigg(1-\frac{\bar{\epsilon}}{2G\Omega}\bigg)~,
\hspace{5mm} v^{2}=\frac{1}{2}\bigg(1+\frac{\bar{\epsilon}}{2G\Omega}\bigg)~,
\label{LNuv}
\end{equation}
where
\begin{equation}
\bar{\epsilon}=\frac{2G\Omega\epsilon}{2G\Omega+\alpha}~,\hspace{8mm} 
\alpha=4\lambda_{2}-G~.
\label{LNepsilon}
\end{equation}
Setting the factor $\alpha=$ 0 recovers the BCS equation without the 
self-energy term, while setting $\lambda_{2}=$ 0 brings back 
to the BCS equation 
including this term.
The factor $\alpha$ is found substituting $(uv)^{2}$ in the equation for $\lambda_{2}$:
\begin{equation}
\frac{4\lambda_{2}}{G}=\frac{\Omega-2(uv)^{2}}{2(2\Omega-1)(uv)^{2}}~.
\label{lambda2}
\end{equation}
This leads 
to the following cubic equation for $\alpha$:
\begin{equation}
\alpha(2\Omega-1)[(\alpha+2G\Omega)^{2}-\epsilon^{2}]-2G\Omega\epsilon^{2}=0~.
\label{LNalpha}
\end{equation}
The ground state energy is given by
\begin{equation}
E_{\rm LN}=\Omega(u^{2}-v^{2})-\frac{\Delta_{\rm LN}^{2}}{G}
-G\Omega[1-2(uv)^{2}]-2(\alpha+G)\Omega(uv)^{2}~.
\label{ELN}
\end{equation}
\begin{figure}[htb]
\begin{center}
\begin{minipage}[t]{150mm}
\epsfxsize=\hsize\epsfbox{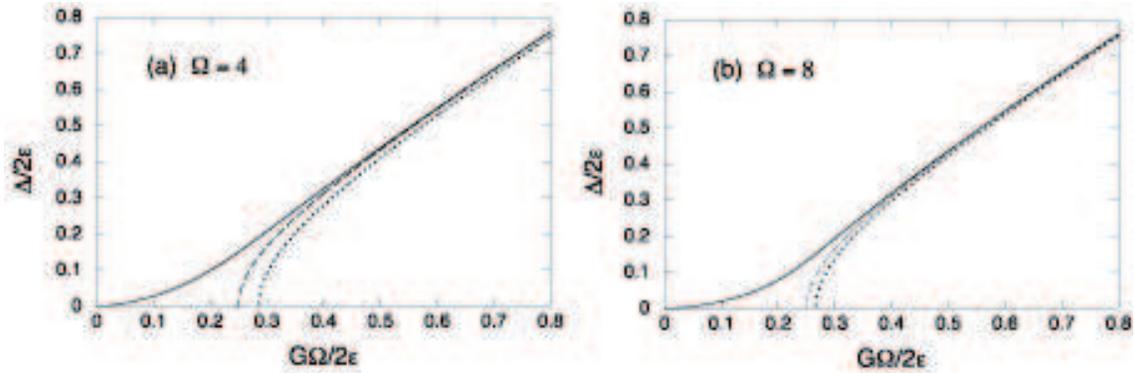}
\end{minipage}
\end{center}
\caption{Pairing gaps (in units of $2\epsilon$) 
as a function of  
$G\Omega/(2\epsilon)$ for $\Omega=$ 4 (a) and 8 (b). 
The dotted line denotes the BCS gap $\Delta$ obtained including 
the self-energy term. The dashed line is the BCS result neglecting 
the self-energy term. 
The solid line stands for the gap $\Delta_{\rm LN}$ given by the LN 
method.\label{fgap}}
\end{figure}
The gaps obtained within the BCS approximation and LN method 
using Eqs. (\ref{gap}) and 
(\ref{LNgap}) for $\Omega=$ 4 and 8 are plotted against the interaction 
parameter $G\Omega/(2\epsilon)$ in Fig. \ref{fgap}.
With decreasing the interaction, the BCS gap $\Delta$ decreases and 
collapses at $G_{\rm cr}$, whose value for the case when the self-energy 
term is neglected is smaller than that obtained including the self-energy 
term. The LN gap $\Delta_{\rm LN}$, 
on the contrary, 
decreases monotonously with decreasing $G$ until $G=0$, where 
$\Delta_{\rm LN}$ vanishes, showing no signature of the superfluid-normal 
phase transition. Hence, by removing the particle number 
fluctuations, the LN method erases completely
the odd behavior of the BCS gap characterized by this phase transition.
The system remains in the superfluid phase at all nonzero values of the 
pairing interaction. It is interesting to see that, 
due to the suppression of the self-energy term $-Gv^{2}$
by $4\lambda_{2}v^{2}$ in the LN method, the BCS gap without the
self-energy correction is closer to the LN result than the BCS gap including 
this correction. By comparing Fig. \ref{fgap} (a) 
and (b), we also see that the 
difference between two BCS versions (with and without the self-energy term) 
decreases significantly when the particle number increases. 
\section{QRPA (RPA), its renormalization, and combination with the LN method}
The RPA ground state includes 
$2p2h$-, $4p4h$-, $6p6h$-, etc. excitations 
on top of
the HF ground state.
The RPA including the pairing correlations within the quasiparticle 
representation is usually referred to as the QRPA.
The correlations in the QRPA ground state, therefore, are much richer
than the $2p2h$-, $4p4h$-, etc. correlations in the ``diffuse" 
(quasiparticle) 
ground state $|0\rangle_{\alpha}$ created by scattering of particle pairs 
within the BCS approximation.
The LN method takes into account some $2p2h$-correlations 
beyond the BCS approximation 
by using the $\lambda_{2}\hat{N}^{2}$ term. However, 
this method still uses the same BCS ground state $|0\rangle_{\alpha}$ since 
it approximately expresses the expectation value of the Hamiltonian with 
respect to the projected state in terms of that with respect to the BCS 
ground state~\cite{LN}.  
The discussion below is conducted within the QRPA, where the RPA is 
obtained as the limit when $\Delta=$ 0. Consequently, we summarize also the 
main features of the renormalized QRPA, whose zero-pairing limit is the 
renormalized RPA. In the present two-level model, the QRPA works at 
$G\geq G_{\rm cr}$, while the RPA is applied
at $G<G_{\rm cr}$.
\subsection{QRPA}
The standard QRPA operators, called phonon operators, have the following 
form in the present two-level model
\begin{equation}
Q^{\dagger}_{\nu}=\sum_{j}(X_{j}^{(\nu)}{\cal A}_{j}^{\dagger}-Y_{j}^{(\nu)}
{\cal A}_{j})~,\hspace{8mm} Q_{\nu}=(Q_{\nu}^{\dagger})^{\dagger}~,
\label{Q}
\end{equation}
The QRPA ground state $|0\rangle_{\rm Q}$ 
is defined as the vacuum for the phonon operator, i.e.
$
Q_{\nu}|0\rangle_{\rm Q}=0={}_{\rm Q}\langle 0|Q^{\dagger}_{\nu}$. 
The $0^{+}$ excited state $|0^{+}\rangle$ 
is obtained by acting $Q^{\dagger}_{\nu}$ on 
this vacuum, i.e.
$
|0^{+}\rangle=Q^{\dagger}_{\nu}|0\rangle_{\rm Q}$. 
The excitation energy $\omega_{\nu}$ 
of the state  $|0^{+}\rangle$, and the amplitudes $X_{j}^{(\nu)}$ and $Y_{j}^{(\nu)}$ 
are found respectively as the eigenenergy and the 
components of the eigenvector of the 
QRPA equation, which is derived from the following 
equation of motion for the Hamiltonian
(\ref{qH}):
\begin{equation}
{}_{Q}\langle 0|[\delta Q,[H,Q^{\dagger}_{\nu}]]|0\rangle_{\rm Q}=
\omega_{\nu}{}_{Q}\langle 0|[\delta Q,Q^{\dagger}_{\nu}]|0\rangle_{\rm Q}~.
\label{eqmotion}
\end{equation}
In the standard way of derivation of the QRPA equations, the BCS equation 
is solved first. Then the a- and b-terms in the Hamiltonian (\ref{qH}) is 
replaced with the BCS result, which is
$H_{\rm BCS}=\sum_{j}E_{j}{\cal N}_{j}$. 
Using the exact commutation relations (\ref{calAA}), 
we see that, among the remaining terms of (\ref{qH}), 
which do not contribute in the BCS, the d-, h-, 
and q-terms start to contribute within the QRPA. The c-term and g-term do not
contribute 
since, in the commutation with the phonon operators (\ref{Q}), 
the former gives a number, while the latter leads to the terms
$\sim {\cal A}^{\dagger}_{j}{\cal A}^{\dagger}_{j'}$,
$\sim {\cal A}^{\dagger}_{j}{\cal A}_{j'}$, and 
${\cal N}_{j}(1-{\cal N}_{j'}/\Omega_{j'})$, which 
are left out by linearizing the equation of motion according to 
(\ref{eqmotion}).
Moreover, in order to obtain a set of QRPA equations, linear with 
respect to the $X_{j}^{(\nu)}$ and $Y_{j}^{(\nu)}$ amplitudes, another  
approximation called the quasiboson approximation is made, which implies
that the following approximate commutation relation
holds 
\begin{equation}
[{\cal A}_{j},{\cal 
A}^{\dagger}_{j'}]=\delta_{jj'}~,
\label{QBA}
\end{equation}
instead of Eq. (\ref{calAA}). The definition of phonon operators 
(\ref{Q}) and the quasiboson approximation (\ref{QBA}) lead
to the well-known 
normalization of the QRPA $X_{j}^{(\nu)}$ and $Y_{j}^{(\nu)}$ amplitudes
\begin{equation}
\sum_{j}[X_{j}^{(\nu)}X_{j}^{(\nu')}-Y_{j}^{(\nu)}Y_{j}^{(\nu')}]=\delta_{\nu\nu'}
\label{XX-YY}
\end{equation}
so that the phonon operators are bosons, i.e.:
\begin{equation}
[Q_{\nu},Q_{\nu'}^{\dagger}]=\delta_{\nu\nu'}
\label{QQ}
\end{equation}
The quasiboson approximation (\ref{QBA}) 
shows that the quasiparticle-pair operators ${\cal A}^{\dagger}_{j}$ and
${\cal A}_{j}$ behave like boson operators when interacting with each 
other. The effect of Pauli principle represented by the last term at the 
RHS of (\ref{calAA}) is just ignored. 
The set of QRPA equations obtained 
in this way is written in the matrix 
form as
\begin{equation}
{~~~A\hspace{8mm} B\choose-B~~-A}{X\choose Y}=\omega{X\choose Y}~.
\label{matrix}
\end{equation}
The explicit form of the matrices $A$ and $B$ depend on the 
approximation. Below we compare the results obtained within the boson and 
fermion formalisms.
\subsubsection{Boson formalism}
The boson formalism is based on two following assumptions:

(a) It considers ${\cal A}_{j}^{\dagger}$ and ${\cal A}_{j'}$ as ideal 
bosons ${\bf b}^{\dagger}_{j}$ and ${\bf b}_{j}$, respectively, 
according to the quasiboson approximation (\ref{QBA}). The 
$j$-shell quasiparticle number operator ${\cal N}_{j}$ is then 
mapped onto a boson pair as
\begin{equation}
{\cal N}_{j}=2{\bf b}^{\dagger}_{j}{\bf b}_{j}.
\label{mappingcalN}
\end{equation}
This mapping preserves the commutators (\ref{calNA}). 
The Hamiltonian (\ref{qH}) can be then fully 
expressed in terms of the boson 
operators ${\bf b}^{\dagger}_{j}$ and ${\bf b}_{j}$. 

(b) In deriving the QRPA equations according to (\ref{eqmotion}), the q-term
of the Hamiltonian (\ref{qH}) is neglected. 
This term is a special case of the so-called 
scattering term in the general 
Hamiltonian with a two-body residual interaction. For instance, when the 
residual interaction is separable, the 
q-term involves a sum of products of two scattering quasiparticle pairs
$\sim B_{jj'}^{\dagger}(LM)B_{j_{1}j_{1}'}(L\widetilde{M})$, 
where $B_{jj'}(LM)=-\sum_{mm'}\langle jmj'm'|LM\rangle
\alpha_{jm}^{\dagger}\alpha_{j'\widetilde{m'}}$
is the scattering quasiparticle-pair operator. The latter is equal to 
${\cal N}_{j}/\sqrt{2j+1}$ when $L=0$. The scattering term is usually omitted in most 
of numerical calculations within QRPA for realistic nuclei in literature.

The boson mapping of the phonon operator (\ref{Q}) becomes
\begin{equation}
Q_{\nu}^{\dagger}\rightarrow \sum_{j}(X_{j}^{(\nu)}{\bf b}_{j}^{\dagger}- 
Y_{j}^{(\nu)}{\bf b}_{j})~.
\label{mappingQ}
\end{equation} 
The QRPA matrices $A_{jj'}$ and $B_{jj'}$  
have the simple form:
\begin{equation}
A_{jj'}=2E_{j}\delta_{jj'}+d_{jj'}~,
\hspace{8mm} 
B_{jj'}=2h_{jj'}~,
\label{ABmatrixboson}
\end{equation}
which, in the present two-level model, become 
\begin{equation}
A_{11}=A_{22}=G\Omega+
\frac{\Delta^{2}}{2G\Omega}~,
\label{A11boson}
\end{equation}
\begin{equation}
A_{12}=A_{21}=-\frac{\Delta^{2}}{2G\Omega}~,
\label{A12boson}
\end{equation}
\begin{equation}
B_{11}=B_{22}=\frac{\Delta^{2}}{2G\Omega}~,
\label{B11boson}
\end{equation}
\begin{equation}
B_{12}=B_{21}=G\Omega-\frac{\Delta^{2}}{2G\Omega}~.
\label{B12boson}
\end{equation}
The QRPA equations give one positive solution equal to 
\begin{equation}
\omega_{\rm QRPA}^{(b)}=2\Delta~,
\label{ombosonQRPA}
\end{equation} 
while the spurious state 
associated with the non-conservation of particle number is located 
exactly at zero energy. We see that the energy of the first excited $0^{+}$ 
state above the phonon ground state $|0\rangle_{\rm Q}$ 
is just twice the pairing gap, the same for the lowest 
two-quasiparticle excitation above the quasiparticle ground state 
$|0\rangle_{\alpha}$ of a system with an even particle number.  
 
In the normal-fluid phase ($\Delta$=0), the 
quasiparticle operator $\alpha_{jm}$ becomes the particle-($p$) creation 
or hole-($h$) destruction operator depending on whether the level is 
located above 
or below the Fermi level, namely
\begin{equation}
\alpha_{jm}^{\dagger}=\left\{\begin{array}{ll}
a_{jm}^{\dagger} & \textrm{if $\epsilon_{j}>\lambda$}\\
-a_{j-m} & \textrm{if $\epsilon_{j}<\lambda$}
\end{array}\right.
\end{equation}
Therefore, the following boson mapping for operators $A_{j}^{\dagger}$ and 
$A_{j}$ holds
\begin{equation}
A_{j}^{\dagger}\rightarrow
\left\{\begin{array}{ll}
{\bf b}^{\dagger}_{j} & \textrm{if $\epsilon_{j}>\lambda$}\\
{\bf b}_{j} & \textrm{if $\epsilon_{j}<\lambda$}
\end{array}\right.~,\hspace{8mm} 
A_{j}\rightarrow\left\{\begin{array}{ll}
{\bf b}_{j} & \textrm{if $\epsilon_{j}>\lambda$}\\
{\bf b}^{\dagger}_{j} & \textrm{if $\epsilon_{j}<\lambda$}
\end{array}\right.~,
\label{mappingA}
\end{equation}
The boson mapping for the number operator $N_{j}$ is given as
\begin{equation}
N_{j}=\left\{\begin{array}{ll}
2{\bf b}_{j}^{\dagger}{\bf b}_{j} 
& \textrm{if $\epsilon_{j}>\lambda$}\\
2(\Omega_{j}-{\bf b}_{j}^{\dagger}{\bf b}_{j}) & \textrm
{if $\epsilon_{j}<\lambda$}
\end{array}\right.~,
\label{mappingN}
\end{equation}
which preserves the commutation relations (\ref{NA}) and 
the particle number on the j-shell equal to $2\Omega_{j}$.
The RPA matrices for the two-level model are
\begin{equation}
A_{11}=A_{22}=2(\epsilon-G\Omega)~,\hspace{8mm} A_{12}=A_{21}=0~.
\label{AbosonRPA}
\end{equation}
\begin{equation}
B_{11}=B_{22}=0~,\hspace{8mm} B_{12}=B_{21}=-2G\Omega~.
\label{BbosonRPA}
\end{equation}
They lead to the RPA phonon energy
\begin{equation}
\omega_{\rm RPA}^{(b)}=2\epsilon\sqrt{1-\frac{G}{G_{\rm cr}^{(b)}}}~,
\hspace{8mm} G_{\rm cr}^{(b)}=\frac{\epsilon}{2\Omega}~.
\label{ombosonRPA}
\end{equation}
The critical value of $G_{\rm cr}^{(b)}$ where
the RPA collapses is the same collapsing point of the BCS 
obtained without the self-energy 
term (Eqs. (\ref{epst}) and (\ref{Gcrit})).
Therefore, in order to have the superfluid regime 
start at the same critical point, the gap $\Delta$ in Eqs. 
(\ref{A11boson}) -- (\ref{ombosonQRPA}) should be calculated 
neglecting the self-energy term 
in the BCS equation (\ref{gap}). This gives
\begin{equation}
\Delta^{(b)}=G\Omega\sqrt{1-\bigg(\frac{\epsilon}{2G\Omega}\bigg)^{2}}~.
\label{Deltab}
\end{equation} 
The phonon energies (\ref{ombosonQRPA}) and (\ref{ombosonRPA}) are exactly 
those obtained for the first time in Ref. \cite{Hogaasen} using
the space-variable technique and the 
gap $\Delta^{(b)}$ (\ref{Deltab}).
\subsubsection{Fermion formalism}
The fermion formalism does not use directly the quasiboson approximation 
(\ref{QBA}). Instead, it employs 
the exact commutation relations (\ref{calAA}) and (\ref{calNA}) 
to rearrange the results of calculating 
the commutators $[H,{\cal A}^{\dagger}_{j}]$ 
and $[H,{\cal A}_{j}]$ into the normal order. 
Then, in the process of linearizing the equation of motion 
(\ref{eqmotion}) 
the following 
``average" quasiboson approximation is used
\begin{equation}
{~}_{\rm Q}\langle 0|[{\cal A}_{j},{\cal A}_{j'}^{\dagger}]|0\rangle_{\rm Q}
=\delta_{jj'}~,
\label{QBAave}
\end{equation}
assuming that the quasiparticle occupation 
number $n_{j}$ in the RPA ground state is 
zero, i.e.
\begin{equation}
n_{j}\equiv\frac{
{}_{\rm Q}\langle 0|{\cal N}_{j}|0\rangle_{\rm Q}}{2\Omega_{j}}=0~.
\label{nj}
\end{equation}

The QRPA matrices are obtained in this way as
\begin{equation}
A_{jj'}=2(E_{j}+2q_{jj})\delta_{jj'}+d_{jj'}~,\hspace{8mm} 
B_{jj'}=2(1-\frac{1}{\Omega_{j}}\delta_{jj'})h_{jj'}~.
\label{matrixAB}
\end{equation}
Their explicit form in the two-level model is given as
\begin{equation}
A_{11}=A_{22}=G\Omega-\frac{\Delta^{2}}{G\Omega^{2}}+
\frac{\Delta^{2}}{2G\Omega}~,
\label{A11}
\end{equation}
\begin{equation}
A_{12}=A_{21}=-\frac{\Delta^{2}}{2G\Omega}~,
\label{A12}
\end{equation}
\begin{equation}
B_{11}=B_{22}=-\frac{\Delta^{2}}{2G\Omega^{2}}+\frac{\Delta^{2}}{2G\Omega}~,
\label{B11}
\end{equation}
\begin{equation}
B_{12}=B_{21}=G\Omega-\frac{\Delta^{2}}{2G\Omega}~.
\label{B12}
\end{equation}
The BCS equation (\ref{gap}) 
for the gap $\Delta$ includes the self-energy term.
The $4q_{jj}\delta_{jj'}$-term in the expression for $A_{jj}$ (Eqs. 
(\ref{matrixAB}) and (\ref{A11})) appears due to the use of the exact 
commutation relation (\ref{calNA}) to calculate the commutator between the 
q-term of the Hamiltonian (\ref{qH}) and ${\cal A}^{\dagger}_{j}$ 
(or ${\cal A}_{j}$) 
as follows:
\begin{equation}
\sum_{j_{1}j_{1}'}q_{j_{1}j_{1}'}[{\cal N}_{j_{1}}{\cal N}_{j_{1}'},{\cal 
A}^{\dagger}_{j}]=4q_{jj}{\cal A}^{\dagger}_{j}+
4\sum_{j'}q_{jj'}{\cal A}^{\dagger}_{j}{\cal N}_{j'}~.
\label{calNNA}
\end{equation}
The last term at the RHS of (\ref{calNNA}) does not contribute in the 
linearization within the QRPA. The first term leads to 
the above-mentioned 4$q_{jj}\delta_{jj'}$-term.
It is worthwhile to notice that, 
if one rearranges the quasiparticle operators in 
the q-term of (\ref{qH}) to the normal order as
\begin{equation}
\sum_{jj'}q_{jj'}{\cal N}_{j}{\cal N}_{j'}=
\sum_{j}q_{jj}{\cal N}_{j}-\sum_{jmj'm'}q_{jj'}\alpha_{jm}^{\dagger}
\alpha_{j'm'}^{\dagger}\alpha_{jm}\alpha_{j'm'}~,
\label{qNN}
\end{equation}
and then drops the last term at the RHS (although there is no 
rigorous justification for doing so), the remaining term in the 
commutation with ${\cal A}^{\dagger}_{j}$ gives
\begin{equation}
\sum_{j_{1}j_{1}'}q_{j_{1}j_{1}'}[{\cal N}_{j_{1}}{\cal N}_{j_{1}'},{\cal 
A}^{\dagger}_{j}]\simeq
\sum_{j'}q_{j'j'}[{\cal N}_{j'},{\cal A}^{\dagger}_{j}]=2q_{jj}{\cal 
A}_{j}^{\dagger}~,
\label{Hagi}
\end{equation}
i.e. just the half of the first term at the RHS of Eq. (\ref{calNNA}). 
This explains the difference between Eq. (\ref{A11}) above and Eq. (38)
of Ref. \cite{Hagino} by a factor of 2 in the denominator of the second 
term at their RHS.

The energy of the first $0^{+}$ excited state  
$\omega$ obtained solving the QRPA equation with 
the matrices (\ref{A11}) -- (\ref{B12}) has the form
\begin{equation}
\omega_{1}=
2\Delta\sqrt{\bigg(1-\frac{3}{4\Omega}\bigg)\bigg
(1-\frac{\Delta^{2}}{4G^{2}\Omega^{3}}\bigg)}\hspace{3mm}~.
\label{omega1}
\end{equation} 
The energy of the spurious state in this case is imaginary.
If one uses the approximation (\ref{Hagi}) instead of (\ref{calNNA}), the 
positive solution is given by Ref. \cite{Hagino} as:
\begin{equation}
\omega_{01}=2\Delta\sqrt{1-\frac{1}{2\Omega}}~,
\label{omega01}
\end{equation}
while the energy of the spurious state is exactly zero.
Neglecting the q-term in (\ref{qH}) yields the energy of the first $0^{+}$ 
state as 
\begin{equation}
\omega_{0}=2\Delta\sqrt{\bigg(1-\frac{1}{4\Omega}\bigg)\bigg
(1+\frac{\Delta^{2}}{4G^{2}\Omega^{3}}\bigg)}~.
\label{omega0}
\end{equation}
There are two features of the fermion formalism, which are worth noticing 
here. The fact that in the process of 
linearizing the equation of motion, the exact commutation relations
(\ref{calAA}) and (\ref{calNA}) have been used to calculate
the commutators $[H,{\cal A}^{\dagger}_{j}]$ and 
$[H, {\cal A}_{j}]$ does not reduce matrices (\ref{matrixAB}) 
by setting $q_{jj'}=$ 0 
to those given within the boson 
formalism (\ref{ABmatrixboson}). The difference in $B_{jj}$ still remains. 
Another feature is that, when the q-term in (\ref{qH}) is omitted, 
the spurious mode in the fermion formalism is shifted to a positive 
value equal to
\begin{equation}
\omega_{0}^{s}=\Delta\sqrt{\frac{1}{\Omega}\bigg(1-\frac{\Delta^{2}}
{4G^{2}\Omega^{3}}\bigg)}~.
\label{omegas}
\end{equation}
However it is
much smaller than the pairing gap $\Delta$ especially 
at large $\Omega$, therefore, remains well-isolated.
Comparing Eqs. (\ref{omega1}) -- (\ref{omega0}) 
with the boson energy $\omega_{\rm QRPA}^{(b)}$ (Eq. 
(\ref{ombosonQRPA})),
it is easy to see that
\begin{equation}
\omega_{1}<\omega_{01}<\omega_{0}<\omega_{\rm QRPA}^{(b)}=2\Delta~.
\label{3omega}
\end{equation}
The ground-state energy is now calculated following Refs. \cite{Rowe,Hagino}
as
\begin{equation}
E_{\rm QRPA}=E_{\rm BCS}+\frac{1}{2}[\omega-(A_{11}+A_{22})]~.
\label{EQRPA}
\end{equation}

In the limit of zero gap, 
the $pp$-RPA equation is obtained from  
the QRPA equation discussed above putting $\Delta=$ 0, $v=v_{1}=u_{2}=$1,  
$u=u_{1}=v_{2}=0$. The solution of this RPA equation is decoupled 
into the addition and removal modes, which have been discussed in 
detail in Ref. \cite{Ring,Hagino}. In the present two-level model, 
these two sets of equations can be written in one matrix equation as
\begin{equation}
{A_{11}\hspace{8mm} B_{12}\choose 
B_{12}\hspace{8mm} A_{22}}{R_{p}^{(\tau)}\choose R_{h}^{(\tau)}}=
{1\hspace{8mm} 0\choose0~~-1}
{R_{p}^{(\tau)}\choose R_{h}^{(\tau)}}\omega_{\tau}~,\hspace{8mm} (\tau=~1,~2~)~,
\label{RPA1RPA2}
\end{equation}
where 
\begin{equation}
A_{11}=\epsilon+2\lambda+2G-G\Omega~,\hspace{8mm} 
A_{22}=\epsilon-2\lambda-G\Omega~,\hspace{8mm} B_{12}=G\Omega~,
\label{RPAmatrix}
\end{equation}
and $R_{p}^{(1)}=X_{\rm a}$~, $R_{h}^{(1)}=Y_{\rm a}$~, 
$R_{p}^{(2)}=Y_{\rm r}$~, $R_{h}^{(2)}=X_{\rm r}$~, 
$\omega_{1}=\omega_{\rm a}$~, and $\omega_{2}=-\omega_{\rm r}$~.
The energies  
$\omega_{\rm a}$ and $\omega_{\rm r}$ are found as
\begin{equation} 
\omega_{\rm a}=-G-2\lambda\pm\sqrt{(\epsilon+G)(\epsilon+G-2G\Omega)}~.
\label{oma}
\end{equation}
\begin{equation}
\omega_{\rm r}=G+2\lambda\pm\sqrt{(\epsilon+G)(\epsilon+G-2G\Omega)}~,
\label{omr}
\end{equation}
where only the sign $+$ in front of the square roots corresponds to the 
positive values for these energies.
Here we still keep the factor $\pm 2\lambda$, which is useful to connect
the RPA solutions with the QRPA one at the critical point $G=G_{\rm cr}$, 
where 2$\lambda=-G_{\rm cr}$ according to Eq. (\ref{lambda}).

In order make a comparison with the boson formalism, where there is only one boson 
state, we apply the sum-rule method, representing 
the RPA phonon operator as
\begin{equation}
Q^{\dagger}=Q_{\rm a}^{\dagger}+Q_{\rm r}^{\dagger}~.
\label{Qa+Qr}
\end{equation}
The RPA ground-state wave function $|\rm{RPA}\rangle$ 
can be written as a direct product of the ground-state wave functions
of the (orthogonal) additional and removal modes
\begin{equation}
|\rm{RPA}\rangle=|0\rangle_{\rm a}\otimes|0\rangle_{\rm r}~.
\label{product}
\end{equation}
Using the Thouless theorem for the energy-weighted sum rule ${\cal S}_{1}$
~\cite{Ring} with respect to the phonon operator $Q^{\dagger}$ (\ref{Qa+Qr})
\begin{equation}
{\cal S}_{1}\equiv\sum_{i}\omega_{\rm RPA}^{(i)}|\langle\nu_{i}|Q^{\dagger}|{\rm 
RPA}\rangle|^{2}=
\frac{1}{2}\langle{\rm HF}|[Q^{\dagger},[H,Q^{\dagger}]]|{\rm HF}\rangle~,
\label{Thouless}
\end{equation}
it is easy to see that the LHS of Eq. (\ref{Thouless}) is equal 
$\omega_{\rm RPA}$ since
$|\nu_{i}\rangle=Q^{\dagger}|{\rm RPA}\rangle$ ($i=$ 1), 
while the RHS is equal to
\[
\frac{1}{2}\bigg[\langle{\rm HF}|[Q_{\rm a}^{\dagger},
[H,Q_{\rm a}^{\dagger}]]|{\rm HF}\rangle
+\langle{\rm HF}|[Q_{\rm r}^{\dagger},
[H,Q_{\rm r}^{\dagger}]]|{\rm HF}\rangle\bigg]=
\]
\begin{equation}
\omega_{\rm a}|\langle\nu_{\rm a}|Q^{\dagger}_{\rm a}|{\rm RPA}\rangle|^{2}
+\omega_{\rm r}|\langle\nu_{\rm r}|Q^{\dagger}_{\rm r}|{\rm RPA}\rangle|^{2}=
\omega_{\rm a}|\langle\nu_{\rm a}|\nu_{\rm a}\rangle|^{2}
+\omega_{\rm r}|\langle\nu_{\rm r}|\nu_{\rm r}\rangle|^{2}=
\omega_{\rm a}
+\omega_{\rm r}~.
\label{oma+omr}
\end{equation}
(The crossing terms 
$\langle{\rm HF}|[Q_{\rm i}^{\dagger},
[H,Q_{\rm i'}^{\dagger}]]|{\rm HF}\rangle$ ($i\neq i'$) vanish as can be 
easily checked using $\langle i|i'\rangle=\delta_{ii'}$).
Therefore
\begin{equation}
\omega_{\rm RPA}=\omega_{\rm a}+\omega_{\rm r}~.
\label{omRPA}
\end{equation}
The ground-state energy is then given by
\begin{equation}
E_{\rm RPA}=E_{\rm HF}+\frac{1}{2}(\omega_{\rm RPA}-A_{11}-A_{22})=
\sqrt{(\epsilon+G)(\epsilon+G-2G\Omega)}-
(\epsilon-G\Omega+G)~,
\label{ERPA}
\end{equation}
which is exactly the same expression 
obtained previously in Eq. (45) of Ref. \cite{Hagino}.

\begin{figure}[htb]
\begin{center}
\begin{minipage}[t]{150mm}
\epsfxsize=\hsize\epsfbox{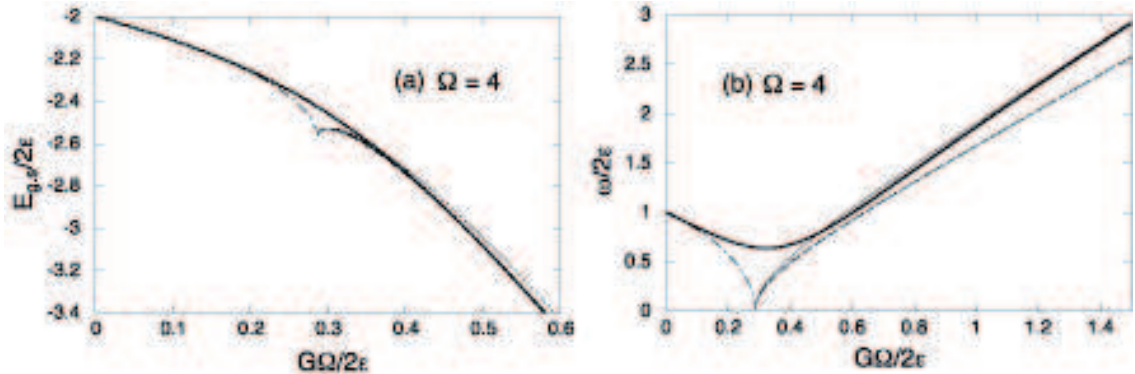}
\end{minipage}
\end{center}
\caption{The ground-state energy (a) and 
energy of the first excited state (b) (in units of $2\epsilon$)
as a function of $G\Omega/(2\epsilon)$.
The thick solid line is the exact result. The thin solid line shows the QRPA 
result neglecting the q-term in the Hamiltonian (\ref{qH}). The dashed 
line denotes the QRPA result of Ref. [6]. The dotted line stands 
for the result obtained using the matrices (\ref{A11}) -- (\ref{B12}).
The dash-dotted line represents the RPA result.\label{fRPA}}
\end{figure}

The energies of the ground-state and the first $0^{+}$ state obtained
within the fermion formalism of RPA and QRPA for $\Omega=$ 4 
are compared with the exact energies in Fig. \ref{fRPA}. 
The figure shows that, in the superfluid regime, 
except for the region close to the critical point 
$G=G_{\rm cr}$ where the QRPA collapses,  
the approximation (\ref{Hagi})
fits the exact result for the ground-state energy 
better than (\ref{calNNA}). 
For the energy of the first $0^{+}$ state, 
both of the QRPA versions, which include the q-term
of the Hamiltonian (\ref{qH}) give practically the same values. 
However, they are obviously smaller compared to the 
exact solution. This discrepancy increases with increasing the 
interaction. From Eqs. (\ref{omega1}) -- (\ref{omega0}) we found that, in 
the limit of infinite $G$ the solution $\omega_{0}$ (q-term neglected) 
becomes $2\Delta\sqrt{1-1/4\Omega}\simeq 3.88\Delta$, while the solutions 
$\omega_{1}$ based on the approximation (\ref{calNNA}), 
and $\omega_{10}$ based on (\ref{Hagi})~\cite{Hagino} are equal to
$0.96\omega_{0}$ and 
$0.93\omega_{0}$, respectively. 
The only QRPA approximation that fits well the exact results for both of the 
ground-state and excited-state energies the one, which neglects the 
q-term of the Hamiltonian (\ref{qH}). The results within this 
approximation 
practically coincide with the exact ones at large $G$.
The RPA energy $\omega_{\rm RPA}=\omega_{\rm a}+\omega_{\rm r}$ 
of the first ${0}^{+}$ state exhibits the well-known 
behavior. It decreases with increasing $G$ and collapses at the same
critical point $G=G_{\rm cr}$, from which the the normal-fluid phase ceases 
to exist, and the superfluid phase begins. Meanwhile, the exact solution 
for the first excited state has only a bending in this region, showing no 
signature of such phase transition.
For the ground-state energy the exact result shows a completely smooth 
curve, while the critical point is clearly seen in the approximations.

\begin{figure}[htb]
\begin{center}
\begin{minipage}[t]{150mm}
\epsfxsize=\hsize\epsfbox{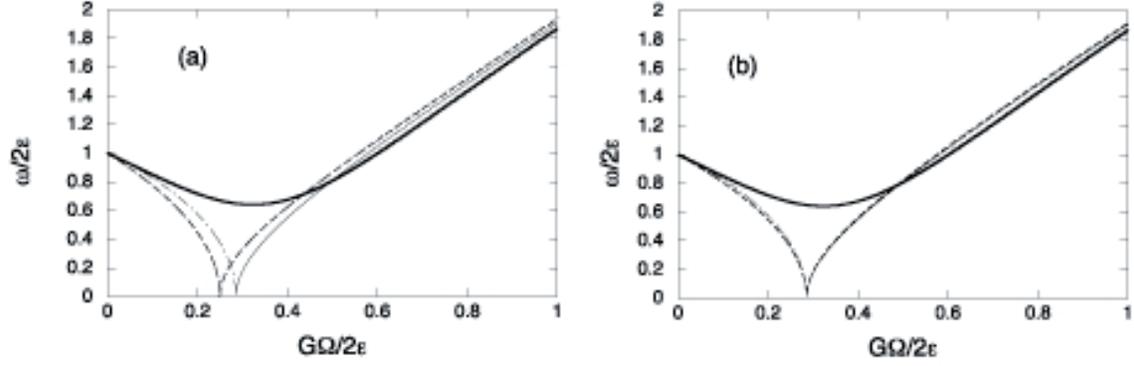}
\end{minipage}
\end{center}
\caption{The energy of the first ${0}^{+}$ state (in units of $2\epsilon$) 
given by the 
boson formalism  (dashed line) in comparison with
those obtained within the QRPA (thin solid line) 
neglecting the q-term of the 
Hamiltonian (\ref{qH}), RPA as the sum $\omega_{\rm a}+\omega_{\rm r}$ 
(dash-dotted line), 
and the exact solution (thick solid line) for $\Omega=$ 4.
In (a) the boson energy [Eqs. (\ref{ombosonQRPA}) and 
(\ref{ombosonRPA})] has been obtained using 
$G_{\rm cr}=G_{\rm cr}^{(b)}=\epsilon/2\Omega$ (neglecting the 
self-energy term). In (b) the boson energy has been calculated using
the value $G_{\rm cr}=\epsilon/(2\Omega-1)$ obtained including the 
self-energy term.  
\label{fboson}}
\end{figure}

As shown in Fig. \ref{fboson} for $\Omega=$ 4, 
the QRPA energy $\omega_{0}$ obtained without 
the q-term of (\ref{qH}), and the RPA energy $\omega_{\rm RPA}$ also 
match well the 
solutions of the boson formalism, 
especially after shifting $G_{\rm cr}=G_{\rm cr}^{(b)}$ 
in the latter to the value of $\epsilon/(2\Omega-1)$ used in the fermion 
formalism due to the inclusion of the self-energy term (Eq. (\ref{Gcrit})). 
\subsection{Renormalized QRPA}
The collapse of the BCS approximation and QRPA (RPA) 
has the same origin of neglecting 
the Pauli principle between quasiparticle pairs operators in the BCS 
approximation (\ref{calA0}) and the quasiboson approximation strictly 
(\ref{QBA}) or in average (\ref{QBAave}).
The LN method approximately corrects this inconsistency within the BCS 
approximation. 
For the QRPA this is done by the renormalized QRPA. 

The essence of the renormalized QRPA is to replace the quasiboson 
approximation in the form of Eqs. (\ref{QBA}) or (\ref{QBAave}) 
with the  average value of the commutator
\begin{equation} 
\langle\widetilde{0}|[{\cal A}_{j},
{\cal A}^{\dagger}_{j'}]|\widetilde{0}\rangle=D_{j}\delta_{jj'}~,
\hspace{8mm} D_{j}=1-2n_{j}~,
\label{beyondQBA}
\end{equation}
in a new ground state $|\widetilde{0}\rangle$, where the correlations
beyond the QRPA due to the fermion structure of the 
quasiparticle pairs ${\cal A}^{\dagger}_{j}$ and ${\cal A}_{j}$ are
taken into account, namely
\begin{equation}
n_{j}=\frac{1}{2\Omega_{j}}\langle\widetilde{0}|{\cal N}_{j}
|\widetilde{0}\rangle\neq 0~,
\label{njQRPA}
\end{equation}
instead of the assumption (\ref{nj}).
The phonon operators are renormalized as
\begin{equation}
{\cal Q}_{\nu}=\sum_{j}\frac{1}
{\sqrt{D_{j}}}({\cal X}_{j}^{(\nu)}{\cal A}^{\dagger}_{j}-{\cal Y}^{(\nu)}_{j}{\cal A}_{j})
~,\hspace{8mm} {\cal Q_{\nu}}=({\cal Q}_{\nu}^{\dagger})^{\dagger}~,
\label{rphonon}
\end{equation}
so that the condition for phonons to be bosons within the correlated ground 
state $|\widetilde{0}\rangle$
\begin{equation}
\langle\widetilde{0}|[{\cal Q}_{\nu},
{\cal Q}^{\dagger}_{\nu'}]|\widetilde{0}\rangle=\delta_{\nu\nu'}~
\label{calQ}
\end{equation}
leads to the same normalization condition for the amplitudes 
${\cal X}_{j}^{(\nu)}$ and ${\cal Y}_{j}^{(\nu)}$ as that of the QRPA, 
i.e. $\sum_{j}({\cal X}_{j}^{(\nu)}{\cal X}_{j}^{(\nu')}
-{\cal Y}_{j}^{(\nu)}{\cal Y}_{j}^{(\nu')})=\delta_{\nu\nu'}~.$
The factor $D_{j}$ is calculated according to the approximation in Ref. 
\cite{RRPA} as
\begin{equation}
D_{j}=\frac{1}{1+({\cal Y}_{j}^{(\nu)})^{2}/{\Omega_{j}}}.
\label{Dj}
\end{equation}
The renormalized-QRPA matrices $A_{jj'}$ and $B_{jj'}$ are given as 
\begin{equation}
A_{jj'}=2(E_{j}+2q_{jj})\delta_{jj'}+
4\sum_{j"}\Omega_{j''}q_{j'j''}(1-D_{j''})+D_{j}d_{jj'}~,
\label{RQRPAmatrixA}
\end{equation}
\begin{equation}
B_{jj'}=2(D_{j}-\frac{1}{\Omega_{j}}\delta_{jj'})h_{jj'}~.
\label{RQRPAmatrixB}
\end{equation}
Since only the omission of the q-term in (\ref{qH}) within the QRPA 
reproduces well the 
exact solution in the present two-level model at large values of $G$, 
we discuss below only the case when $q_{jj'}=$0. The 
renormalized QRPA matrices in this case read
\begin{equation}
A_{11}=A_{22}=G\Omega(2-D)+\frac{D\Delta^{2}}{2G\Omega^{2}}~,
\hspace{8mm} A_{12}=A_{21}=-\frac{\Delta^{2}}{2G\Omega}(4-3D)~,
\label{RQRPA-A}
\end{equation}
\begin{equation}
B_{11}=B_{22}=-\frac{\Delta^{2}}{2G\Omega^{2}}+\frac{D\Delta^{2}}{2G\Omega}~,
\hspace{8mm} B_{12}=B_{21}=D(G\Omega-\frac{\Delta^{2}}{2G\Omega})~.
\label{RQRPA-B}
\end{equation} 
The energy of the first excited $0^{+}$ state is found as
\begin{equation}
\omega_{\rm RQRPA}=
2\Delta\sqrt{\bigg[\bigg(D-\frac{1}{4\Omega}\bigg)+\frac{G^{2}\Omega^{2}(1-D)}
{\Delta^{2}}\bigg]\bigg(1+\frac{\Delta^{2}}{4G^{2}\Omega^{3}}\bigg)}~,
\label{omRQRPA}
\end{equation}
By setting $D_{j}=$ 1 ($n_{j}=$ 0) 
in Eqs. (\ref{RQRPA-A}) -- (\ref{omRQRPA}), 
the QRPA limit in Eqs. (\ref{A11}) -- (\ref{B12}) (without the second term 
at the RHS of (\ref{A11}) since the q-term is neglected), and $\omega_{0}$ given by 
Eq. (\ref{omega0}) are recovered .
The ground-state energy is calculated using Eq. (\ref{EQRPA}), using
the energy $\omega_{\rm RQRPA}$ given by Eq. (\ref{omRQRPA}) and
the matrices $A_{11}=A_{22}$ given by Eq. (\ref{RQRPA-A}).

The renormalized RPA matrices ($G<G_{\rm cr}$) are given as
\begin{equation}
A_{11}=\epsilon+2G-DG\Omega~,\hspace{8mm} 
A_{22}=\epsilon-DG\Omega~,\hspace{8mm} A_{12}=A_{21}=0~,
\label{RRPA_A}
\end{equation}
\begin{equation}
B_{11}=B_{22}=0~,\hspace{8mm} B_{12}=B_{21}=DG\Omega~.
\label{RRPA_B}
\end{equation}
The positive renormalized RPA phonon energies are found as
\begin{equation}
\widetilde{\omega}_{\rm a}
=-G-2\lambda+\sqrt{(\epsilon+G)(\epsilon+G-2DG\Omega)}~.
\label{roma}
\end{equation}
\begin{equation}
\widetilde{\omega}_{\rm r}=G+2\lambda+\sqrt{(\epsilon+G)(\epsilon+G-2D
G\Omega)}~,
\label{romr}
\end{equation}
\begin{equation}
\omega_{\rm RRPA}=\widetilde{\omega}_{\rm a}+\widetilde{\omega}_{\rm r}=
2\sqrt{(\epsilon+G)(\epsilon+G-2DG\Omega)}~,
\label{omRRPA}
\end{equation}
and the ground-state energy is
\begin{equation}
E_{\rm 
RRPA}=\sqrt{(\epsilon+G)(\epsilon+G-2DG\Omega)}-(\epsilon+G-DG\Omega)~.
\label{ERRPA}
\end{equation}
As has been shown in Ref. \cite{RRPA}, the presence of the factor $D$ in the
renormalized RPA matrices reduces the interaction in such a way 
that the critical point where the RPA collapses is completely 
washed out. The energy $\widetilde\omega_{\rm a}$, 
$\widetilde\omega_{\rm r}$, and consequently, $\omega_{\rm RRPA}$ are 
always real. As for the renormalized QRPA, the presence of the pairing 
gap $\Delta$ makes it still collapse if $\Delta$ is calculated within 
the BCS (Eq. (\ref{gap}), 
but it is no longer the case if the LN pairing gap $\Delta_{\rm LN}$ 
(\ref{LNgap}) is used.
\subsection{LN-method + renormalized RPA (renormalized QRPA)}
\label{approx}
We have seen in the preceding sections that the LN-method allows us to 
avoid the phase transition of the BCS approximation, the QRPA without the 
q-term in (\ref{qH}) provides us with the best fit of the exact results
for the energies of the ground state and the first excited ${0}^{+}$ 
state, while the renormalized 
RPA is known to smear out the collapse of the RPA due to the violation of
the Pauli principle within the quasiboson approximation~\cite{RRPA}.
Therefore, in order to avoid the phase transition and to give at the same time a good 
description for both of the energies of the ground state as well as the fist 
excited ${0}^{\dagger}$ state, we propose here the following recipe. 
\begin{figure}[htb]
\begin{center}
\begin{minipage}[t]{150mm}
\epsfxsize=\hsize\epsfbox{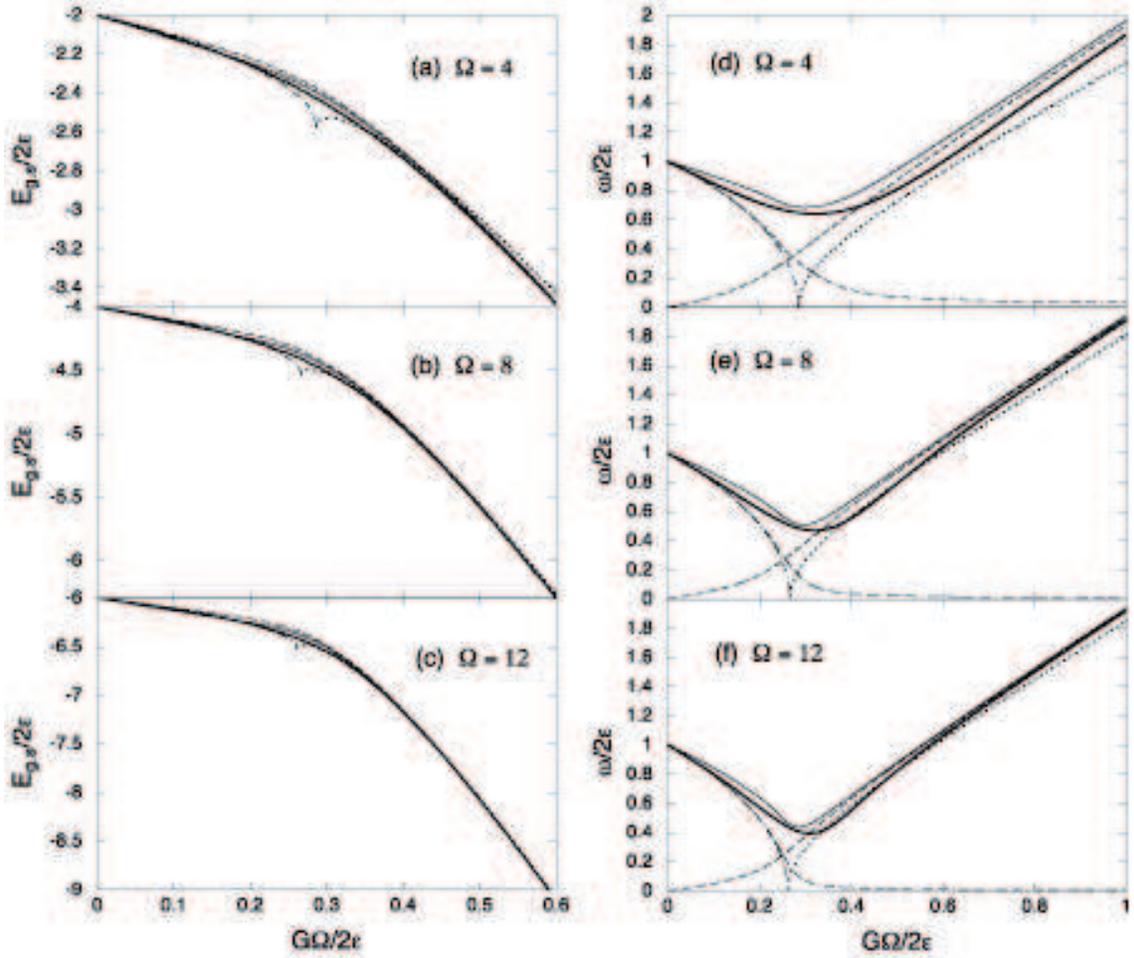}
\end{minipage}
\end{center}
\caption{The energies of the ground state (left panels) and
of the first ${0}^{+}$ excited state (right panels) (in units of $2\epsilon$) 
at several values of $\Omega$. 
The thick solid line is the exact result.
The dotted line is the QRPA result using the matrices (\ref{A11}) -- 
(\ref{B12}). The dash-dotted line denotes the RPA 
result. In (a) -- (c): the dashed line shows the LN result, while the thin solid 
line represents the ground-state energy calculated according to (ii) of 
Section \ref{approx}.
In (d) -- (f): the dashed line stands for renormalized QRPA result using the 
LN pairing gap and neglecting the q-term in (\ref{qH}); 
the double-dash-dotted line represents the renormalized RPA results;
the thin solid line is the energy $\omega_{0^{+}}$ according to
Eq. (\ref{ommixed}).
\label{fECom}}
\end{figure}

i) The QRPA equations (without the q-term in 
(\ref{qH})) are solved using the pairing 
gap $\Delta_{\rm LN}$ (\ref{LNgap}) found in the LN method.

ii) The ground state energy is calculated using Eq. (\ref{EQRPA}), in which
the LN pairing gap $\Delta_{\rm LN}$ is used instead of the BCS gap $\Delta$
(\ref{gap}) and the renormalized QRPA matrices given by Eq. (\ref{RQRPA-A})  
are used instead of Eq. (\ref{A11}).

iii) The first excited $0^{+}$ state 
is presented as a mixed state of those obtained within 
the renormalized RPA and renormalized QRPA (in (i)). Applying the sum-rule 
method and following a derivation similar to Eqs. 
(\ref{product}) -- (\ref{oma+omr}), we find that the
energy $\omega_{0^{+}}$ of this mixed state is the sum of the renormalized
RPA and renormalized QRPA energies, namely
\begin{equation}    
\omega_{0^{+}}=\omega_{\rm RRPA}+\omega_{\rm RQRPA}~,
\label{ommixed}
\end{equation}
where $\omega_{\rm RRPA}$ and $\omega_{\rm QRPA}$ are given by Eqs. 
(\ref{omRRPA}) and (\ref{omRQRPA}), respectively, with $\Delta_{\rm LN}$ 
used in the latter.
The results obtained using this recipe will be referred to as the combined 
results.

The combined results for 
ground state energy and the energy of the first excited $0^{+}$ state
are presented as thin solid lines
in Fig. \ref{fECom} at several values of $\Omega$ in comparison with the 
exact results and those of RPA, QRPA and their renormalized versions.
The result obtained within the renormalized RPA and the combined result (i)
for the energy of the first excited $0^{+}$ state show that the 
superfluid-normal phase
transition is completely washed out. 
For the ground-state energy, the combined result (ii) is closer to the
exact one compared to the LN result in the 
weak-coupling region ($G\ll G_{\rm cr}$) 
and in the region around $G_{\rm cr}$. 
It coincides with the LN and exact results in the strong-coupling
region ($G\gg G_{\rm cr}$) (See the thin lines in Fig. \ref{fECom} (a) -- 
(c)). 
The combined result (iii) agrees
reasonable well with the exact results (See the thin lines in Fig. 
\ref{fECom} (d) -- (f)), especially for larger $\Omega$. 
Hence, this numerical test confirms the validity of the
assumption (iii) that 
the exact excited $0^{+}$ state may be considered as a mixed state
where both of the normal and superfluid phases coexist at $G\neq$ 0. 
The weak-coupling region is dominated by the 
normal-fluid phase, where the energy $\omega_{0^{+}}$ of the first excited $0^{+}$ state 
decreases monotonously with increasing the pairing interaction $G$.  
The strong-coupling region
is dominated by the superfluid phase, in which $\omega_{0^{+}}$ increases 
with increasing $G$. 
\section{Conclusions}
In this work, a schematic model of two 
symmetric levels interacting via a pairing force 
has been used to test several well-known approximations
by calculating the ground-state energy and the energy $\omega_{0^{+}}$ of 
the first excited $0^{+}$ state. Results obtained 
within the BCS approximation, 
LN method, boson and fermion formalisms for RPA and QRPA, 
renormalized RPA, and renormalized QRPA have been 
analyzed in details. It is shown that the common version of the QRPA, 
which neglects the scattering term (the q-term of the Hamiltonian 
(\ref{qH})) and the boson formalism give 
the closest results to the exact ones for both of the 
ground-state energy and $\omega_{0^{+}}$ at the values of 
the pairing interaction parameter $G\gg G_{\rm cr}$. 
Meanwhile 
the QRPA version used in Ref. \cite{Hagino} as well as the QRPA using the 
exact commutation relations between quasiparticle-pair operators to treat 
the scattering term do not describe well the energy of the first excited 
$0^{+}$ state at large $G$.

A recipe has been proposed which combines the results obtained within the LN 
method for the pairing gap $\Delta_{\rm LN}$, 
the renormalized RPA, and the renormalized QRPA neglecting the 
last term (q-term) of the Hamiltonian (\ref{qH}).
The combined results agree reasonably well with the exact ones for both
of the ground-state energy and $\omega_{0^{+}}$, showing no signature of 
a sharp superfluid-normal phase transition. The agreement is better at 
a larger particle number. The results suggest that the exact excited 
$0^{+}$ state can be decomposed approximately into two components, which 
correspond to the normal and superfluid regimes, respectively.
The weak-coupling region is dominated by the normal-fluid phase, while the
strong-coupling region - by the superfluid phase. 
Since the proposed scheme is based on rather simple, well-known, and 
numerically accessible approximations, its future 
extension toward an application in realistic nuclei may be useful.

For a more self-consistent approach, one can derive the set of RPA 
equations using the Hamiltonian (\ref{qH}) with the coefficients $u_{j}$, 
$v_{j}$, $X_{j}^{(\nu)}$ and $Y_{j}^{(\nu)}$ left as variational parameters 
when minimizing the average energies over the ground state and the first 
excited $0^{+}$ state. This may serve as a goal of the future study.

\section*{Acknowledgments}
Thanks are due to V. Zelevinsky for reading the manuscript, valuable 
comments and suggestions.

\end{document}